\documentclass[reprint,superscriptaddress,amssymb,amsmath,aps,showpacs,10pt,floatfix,prl,longbibliography]{revtex4-1}

\usepackage{graphicx}
\usepackage{color}
\usepackage{lipsum}
\usepackage{marvosym}
\usepackage{xcolor}
\usepackage{stackengine}

\usepackage{blindtext}
\usepackage{epstopdf}
\usepackage{amssymb}
\usepackage{amsmath}
\usepackage{amsfonts}
\usepackage[note-name=, use-sort-key = false]{notes2bib}
\usepackage{color}
\usepackage[colorlinks,bookmarks=false,citecolor=darkblue,linkcolor=red,urlcolor=blue]{hyperref} 
\definecolor{darkred}{rgb}{0.7,0.0,0.0}

\definecolor{darkblue}{rgb}{0,0.02,0.45}

\definecolor{darkgreen}{rgb}{0.02,0.45,0.0}

\definecolor{violet}{rgb}{0.8,0.2,0.6}

\begin{document}
\title{Spin-reorientation-induced band gap in Fe$_3$Sn$_2$: Optical signatures of Weyl nodes}

\author{A. Biswas}
\affiliation{1. Physikalisches Institut, Universit{\"a}t Stuttgart, 70569 Stuttgart, Germany}

\author{O. Iakutkina}
\affiliation{1. Physikalisches Institut, Universit{\"a}t Stuttgart, 70569 Stuttgart, Germany}

\author{Q. Wang}
\affiliation{Department of Physics and Beijing Key Laboratory of Opto-electronic Functional Materials \& Micro-nano Devices, Renmin University of China, Beijing 100872, China}

\author{H. C. Lei}
\email{hlei@ruc.edu.cn}
\affiliation{Department of Physics and Beijing Key Laboratory of Opto-electronic Functional Materials \& Micro-nano Devices, Renmin University of China, Beijing 100872, China}

\author{M. Dressel}
\affiliation{1. Physikalisches Institut, Universit{\"a}t Stuttgart, 70569 Stuttgart, Germany}

\author{E. Uykur}
\email{ece.uykur@pi1.physik.uni-stuttgart.de}
\affiliation{1. Physikalisches Institut, Universit{\"a}t Stuttgart, 70569 Stuttgart, Germany}

\date{\today}

\begin{abstract}
Temperature- and frequency-dependent infrared spectroscopy identifies two contributions
to the electronic properties of the magnetic kagome metal Fe$_3$Sn$_2$: two-dimensional Dirac fermions and  strongly correlated flat bands.
The interband transitions within the linearly dispersing Dirac bands appear as a two-step feature along with a very narrow Drude component due to intraband contribution.
Low-lying absorption features indicate flat bands with multiple van Hove singularities.
Localized charge carriers are seen as a Drude-peak shifted to finite frequencies. The spectral weight is redistributed when the spins are reoriented at low temperatures; a sharp mode appears suggesting the opening of a gap due to the spin reorientation as the sign of additional Weyl nodes in the system.
\end{abstract}

\maketitle

Magnetic kagome metals are emerging as a new class of materials with special crystal structures,
which are supposed to bring together electronic correlations, magnetism, and topological orders \cite{Liu2019}.
Merging the strong electronic correlations with the topologically nontrivial states
makes new types of exotic phenomena possible ranging from fractional quantum Hall effect to axion insulators.

Unfortunately, the realization of the materials in this regard is scarce \cite{Nakatsuji2015, Nayak2016, Liu2018, Xu2018}. The FeSn-binary compounds are possible candidates;
Fe$_3$Sn$_2$\ is one of them, where the linearly dispersing Dirac bands lying below the Fermi energy
are confirmed as well as flat bands around $E_F$.
The crystal structure of Fe$_3$Sn$_2$\ protects the inversion and three-fold rotational symmetry,
while the time reversal symmetry is broken due to the magnetic nature of the system.
The unique structural properties of this system give rise the flat-band ferromagnetism \cite{Lin2018}, anomalous Hall effect \cite{Wang2016, Li2019}, and topological Dirac states \cite{Ye2018}.
Due to the strong influence of magnetism, a large tunability of the spin-orbit coupling \cite{Yin2018} and of the massive Dirac fermions \cite{Ye2018, Lin2020} was proposed as well as the emergence of further Weyl nodes at the Fermi energy \cite{Yao2018}.

Fe$_3$Sn$_2$\ consist of bilayer kagome network separated via stanene layers.
The bilayer structure gives rise to the interlayer hybridization due to the multiple $d$ orbitals of Fe atoms leading to deviations from the ideal single-orbital two-dimensional kagome lattice scenario \cite{Mielke1991, Mielke1991a, Mielke1992}. For instance, the flat bands do not extend over the entire Brillouin zone and show a small dispersion \cite{Lin2018}; moreover, correlations among the Dirac bands open a gap at the crossing points that give rise to the correlated massive Dirac fermions \cite{Ye2018}. Despite these deviations from the ideal scenario, this system provides a beautiful playground for investigating the interplay between magnetism, strong electronic correlations, and topological orders.

For Fe$_3$Sn$_2$\ it is known that even in the absence of an external magnetic field the spins reorient at reduced temperatures; despite the fact that the system orders ferromagnetically at much higher $T \approx 640$~K \cite{Kumar2019}. The implications of this reorientation on the electronic structure remain an open question. Considering the large sensitivity of magnetism on temperature, here we employ tem\-per\-a\-ture-dependent broadband infrared spectroscopy for studying this model compound. We look for optical fingerprints of Dirac fermions, localized electrons of the flat bands, and spin reorientation, as they directly probe the interplay between these unique states along with the energy scales. In the absence of an external magnetic field, we investigate the effects of the inherent magnetization on the observed properties. Moreover, the high spectral resolution of our technique
even in the low energy range, gives us the opportunity to test theoretical proposals regarding the additional Weyl nodes in the vicinity of $E_F$.


Single crystals of Fe$_3$Sn$_2$\ were synthesized as described elsewhere \cite{Wang2016}. Here we performed temperature-dependent optical reflectivity measurements on single crystals of Fe$_3$Sn$_2$\ in a broad energy range. Single crystals of Fe$_3$Sn$_2$\ were grown using self-flux method as described elsewhere \cite{Wang2016}. An as-grown sample with a good surface quality was chosen for the optical spectroscopy study. The (001)-plane lateral dimensions of the sample used in the infrared spectroscopy study is 1000~$\mu$m$ \times $800~$\mu$m$ \times$ 200~$\mu$m.

\begin{figure}
\centering
\includegraphics[width=1\linewidth]{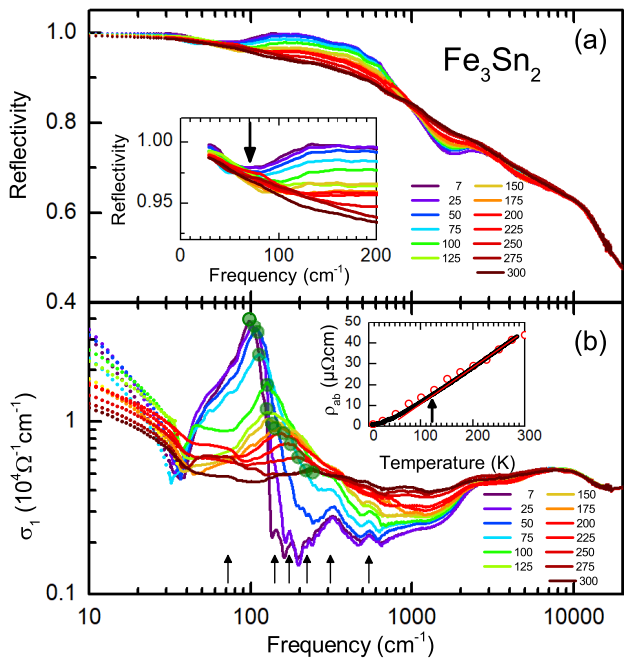}%
\caption{(a) Temperature-dependent reflectivity of Fe$_3$Sn$_2$\ in a broad frequency range. The inset magnifies  the low-energy reflectivity where the dip causes the peak structure in $\sigma_1(\omega)$. (b) Corresponding optical conductivity at different temperatures.
The green circles indicate the peak position. The arrows mark the points of the van Hove singularities. The inset displays the measured dc resistivity of the sample.
RRR = 39 indicates the good quality of our sample. The red circles are the dc resistivity values determined via the optical parameters matches very well with the measured resistivity.  }%
\label{RefOC}%
\end{figure}

Fig.~\ref{RefOC} displays the temperature-dependent reflectivity along with the optical conductivity 
in the entire measured range. Consistent with the highly metallic nature of the sample, the low-energy reflectivity reaches almost unity and the optical conductivity approaches the dc conductivity values on the order of $10^5~(\Omega {\rm cm})^{-1}$ at low $T$. While the optical properties are basically  temperature-independent above approximately 1~eV, a series of interesting features are identified below this energy:
(i)~A strong suppression of the reflectivity and, concomitantly, the optical conductivity starting from the mid-infrared range.
(ii)~A peak-like structure (marked with green circles) that shifts to lower energies with decreasing $T$.
(iii)~A very narrow Drude component that gets even sharper upon cooling.
The scattering of this Drude component is very small and barely visible in our measurement window; however, the dc resistivity data [the inset of the Fig.~\ref{RefOC}(b)] corroborate its existence \bibnote{It is interesting to examine the scattering rate of this sharp Drude-component. As the momentum-relaxing scattering of the free carriers (in the current case is the Dirac fermions as explained in the text) can be observed in the optical conductivity. Hence, one can directly obtain the corresponding scattering time, $\tau$, from the real part of the optical conductivity, $\sigma_1(\omega)$. The simultaneous fits of the reflectivity and the optical conductivity of our low energy Drude-component reveals a scattering rate [$\gamma = 1/(2\pi c\tau)$] of 2~cm$^{-1}$\ at 7~K, while it increases to $\sim$17~cm$^{-1}$\ at room temperature. These corresponds to the scattering times $\tau_ {7K} = 2.5$~ps and $\tau_ {300K} = 0.3$~ps. Considering the reported average Fermi velocity for Fe$_3$Sn$_2$, $v_F = 2.2\times10^5$~m/s, the momentum relaxation lengths [$\ell = v_F\tau$] are calculated as $\ell_ {7K} = 0.55$~$\mu$m and $\ell_ {300K} = 66$~nm. Considering that the localized carrier response of the flat bands are separated in energy and does not contribute to the low energy dynamics (within the obtained scattering scales), these length scales suggest that Fe$_3$Sn$_2$\ readily might be a suitable platform to realize a ballistic transport.}.

\begin{figure}
\centering
\includegraphics[width=1\linewidth]{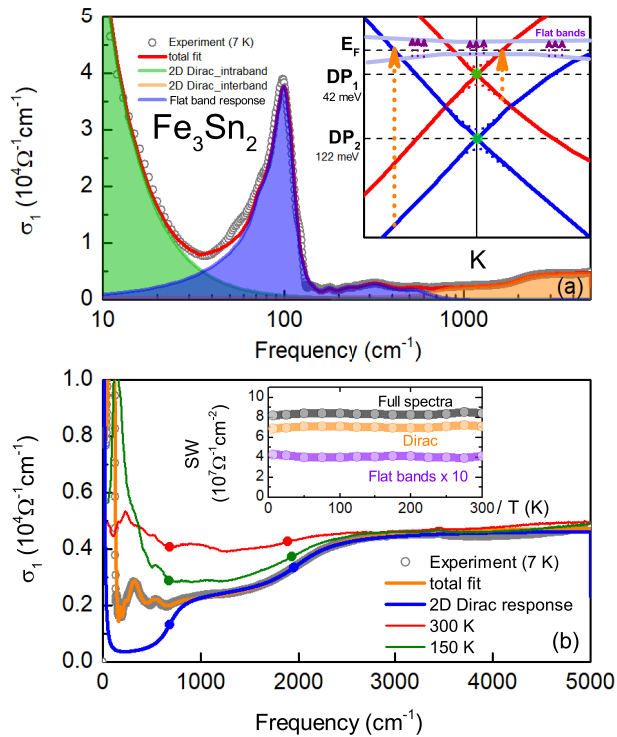}%
\caption{(a) Decomposition of the 7~K optical conductivity. The red curve represents the fit of the overall spectrum with a Drude (green), flat band responses (blue) and two-dimensional Dirac responses (orange). The inset sketches the band structure along with the transitions. (b) Interband and intraband responses of the two-dimensional Dirac fermions (blue curve) after the flat band contributions have been subtracted from the overall fit. The two step feature and the Dirac points are highlighted. Signs of these points are visible at higher $T$ and do not change significantly with temperature; the $T=150$ and 300~K spectra are given for comparison. The inset shows the temperature-dependent SW  analysis. }%
\label{Decomposition}%
\end{figure}

We decomposed the optical conductivity into two main parts shown in Fig.~\ref{Decomposition}(a).
While the high-energy absorption and the low-energy Drude-component
can be interpreted within the Dirac fermions framework,
the strong absorption features in between have to be treated separately.
We attribute these features to the charge carriers within the flat bands, as explained later. This decomposition is supported by the analysis of the spectral weight (SW) [inset of Fig.~\ref{Decomposition}(b)].
The transfer of SW takes place within the different contributions; the overall SW, but also the Dirac and the flat-band spectral weights are conserved.

But first let us discuss the Dirac physics and its optical signatures in Fe$_3$Sn$_2$.
The results of angle resolved photoemmision spectroscopy (ARPES) \cite{Ye2018, Yao2018}, scanning tunneling microscopy (STM) \cite{Lin2018}, and magneto-transport measurements \cite{Wang2016, Liu2018, Li2019, Ye2018, Kumar2019} evidence the linearly dispersing bands of massless Dirac fermions and flat bands of massive localized electrons.
ARPES data indicate that the Dirac bands lie well below $E_F$ (within $\sim$0.15~eV); whereas the magneto-transport measurements also verify the existence of a topologically nontrivial state. Moreover, the underlying bilayer structure of the kagome lattice gives rise to correlations among the Dirac fermions causing a correlation gap to open at the Dirac points.

The optical signatures of these Dirac points are clearly visible in our spectra.
The mid-infrared absorption and the accompanying low-energy Drude component can be well reproduced by taking into account the intraband and interband responses of two-dimensional Dirac fermions as shown in graphene, for instance \cite{Li2008, Scharf2013}.
For a two-dimensional Dirac system with the Dirac point at the Fermi energy and in the absence of other contributions, the optical conductivity is expected to exhibit a frequency-independent behavior.
On the other hand, the shift of the Dirac point with respect to $E_F$
results in absorption feature, where the SW is transferred to the intraband response of the Dirac bands. This situation forbids low energy transitions (up to $2E$, where $E$ defines the energy shift between $E_F$ and the Dirac point) and the optical conductivity starts to increase above a certain energy that is defined as the Pauli edge. In the two-dimensional case, one expects a step-like increase starting from $2E$ and  leveling off to the $\omega$-independent behavior  \cite{Gusynin2006}.

In Fig.~\ref{Decomposition}(b), the 7~K spectrum is plotted with the overall fit to our data,
and after subtracting the low-lying absorption features. This remaining part represents the inter- and intraband contributions of the Dirac fermions; the resulting blue curve reproduces the high-energy interband transitions very well.
In our case, these high-energy intraband transitions involve not one step-like feature,
but actually two. Hence, a somewhat more complicated picture is present in this system and the response of the Dirac fermions cannot be explained within a single Dirac cone picture.
This conclusion is in line with ARPES results \cite{Ye2018},
as depicted by the two-cone picture in Fig.~\ref{Decomposition}(a).
In turn, this gives rise to the two-step absorption feature of the optical conductivity.
We also like to point out that even the high-temperature spectra reveal signatures of these two step absorption feature. In Fig.~\ref{Decomposition}(b), the $T=300$ and 150~K spectra are given
for comparison; the steps due to the Dirac bands are marked by colored dots.

From our spectrum at $T=7$~K we estimate the positions of the Dirac points at $E_{D1}$~=~346~cm$^{-1}$\ (42~meV) and $E_{D2}$~=~983~cm$^{-1}$\ (122~meV); they do not change significantly with temperature.
These values are well in line with ARPES measurements \cite{Ye2018}, while they do not consider the correlation gap observed in ARPES.
Such an energy gap due to correlated Dirac fermions
resembles the optical response known from  density-wave systems \cite{Uykur2019}.
However, in the present case we should not see such an effect, as the gap is buried well below $E_F$. 

Let us turn to the low-lying absorption band of the spectra. We associate these features with the response of the flat bands, since  they are located in the vicinity of the Fermi energy.
We can identify several absorption features in the spectra at 72, 141, 172, 223, 311, 542~cm$^{-1}$; however, a close look reveals that most of them do not shift with decreasing temperature, but simply get sharper [marked by the arrows in Fig.~\ref{RefOC}(b)]. We attribute these peaks to van Hove singularities of the flat bands. As shown by STM measurements \cite{Lin2018}, flat bands do not extend over the entire Brillouin zone. They possess a small dispersion giving rise to several peaks in the density of states around the Fermi energy. Due to the band dispersion, the transitions across the Fermi energy between these flat bands occur at slightly different energies, as observed.
 
\begin{figure}
\centering
\includegraphics[width=1\linewidth]{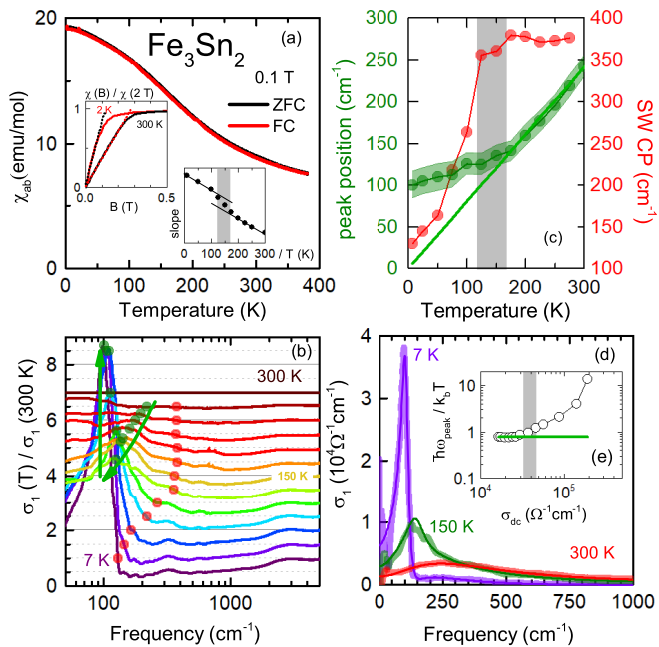}%
\caption{(a) Magnetic susceptibility of Fe$_3$Sn$_2$\ measured for $H\parallel ab$-plane in field-cooled and zero-field-cooled configuration. The inset shows the field dependence $\chi(B)$ normalized to 2~T value (above the saturation field) for $T=300$ and 2~K. Also shown is the slope of the low-field magnetization as a function of $T$ indicating the spin-reorientation transition.
(b) Relative optical conductivity normalized to the room temperature spectrum. 
The spectra for $T>7$~K are shifted by 0.5 each with respect to each other.
The red circles mark the point where the spectra cross each other, indicating the spectral weight transfer energy. The green circles are the maximum of the low-lying absorption feature shifting in energy. Green arrows are guide to the eye following the decreasing temperature. 
Panel (c) shows the $T$-dependence of the red and green circles in panel (b). The solid green line represent the position of the expected localization peak extended to the lower temperatures, highlighting the clear deviation. The same scaling also given in panel (e). (d) Energy-dependent absorption feature at $T=300$, 150, and 7~K with the localization peak fit from reference ~\cite{Fratini2014}. (e) Scaling of the peak position with the dc conductivity suggested by  \cite{Luca2017}. Grey shaded areas in figures represent the crossover temperature, where the spin reorientation takes place. }%
\label{Magnetic}%
\end{figure}

 The absorption feature around 100~cm$^{-1}$\ is by far the most prominent one with a distinct dynamics:
it changes its shape and strongly shifts in energy upon cooling.
Our detailed analysis reveals that this feature is strongly linked to the underlying magnetic structure. Fe$_3$Sn$_2$\ possesses flat-band ferromagnetism related to the underlying kagome lattice with a ferromagnetic phase transition at $T_C=640$~K. Previous magnetization, neutron diffraction, and Mössbauer spectroscopy studies all conclude that the spins are canted towards out of plane at high temperatures up to $T_C$.
With decreasing temperature they tend to reorient towards the kagome plane \cite{Trumpy1970, Caer1978, Malaman1978, Caer1979}. However, no agreement has been reached on the temperature range, where this reorientation occurs, and whether the phase transition is of first or a second order. A recent magneto-transport study showed that the spin reorientation takes place in a temperature range between 70-150~K with a transition peak at $T=120$~K \cite{Kumar2019}. The nature of this spin reorientation and its implications on the electronic structure remain to be clarified. Characterizing our sample in this regard, in Fig.~\ref{Magnetic}(a) we plot the magnetic susceptibility as a function of temperature. Our findings are consistent with the literature and yield a crossover temperature slightly below 150~K.

Let us turn back to the strong optical absorption around 100~cm$^{-1}$.
To better analyse the evolution of the peak, we plotted the relative optical conductivity [$\sigma_1(T)/\sigma_1(300~K)$] in Fig.~\ref{Magnetic}(b), where one can trace the energy position of the peak as well as the transfer of spectral-weight to the low-lying absorption features. Although we cannot rule out that appear accidentally, at elevated temperatures a clear isosbestic point (indicated by the red symbols, crossing of the spectra at 300~K spectrum, see Supplementary Materials for the details) of the spectra can be defined along with the SW transfer that is lead by the temperature change \cite{Greger2013}. Below the spin reorientation temperature the isosbestic behavior does not hold anymore and the spectra crossing point rapidly shifts to the smaller energies, as the SW accumulates to a very sharp peak structure. From panel (c) we can see that its maximum (green circles) gradually moves to lower energies: while for $T>150$~K it decreases linearly in $T$, the shift tends to saturate at lower temperatures. The drastic change of the isosbestic signature suggests the influence of another mechanism other than the temperature on the SW redistribution.

Since the flat bands possess strongly correlated, localized charge carriers, the so-called localization peak is a plausible assumption for the observed peak. The generalization of the Drude response is commonly discussed in the framework of the strongly correlated electron systems \cite{Smith2001, Fratini2014, Luca2017}. The partial localization of the charge carriers leads to a displaced Drude-peak, where the low-energy part of the optical conductivity is strongly suppressed and the peak-like structure appears at finite energies. Literature is rich in this regard, with examples from numerous transition metal oxides \cite{Kostic1998, Lee2002, Bernhard2004, Wang2004, Puchkov1995, *Hwang2007, Tsvetkov1997,  Osafune1999, *Uykur2011, Rozenberg1995, Joensson2007, Takenaka1999, *Takenaka2002, Jaramillo2014} and organic conductors \cite{Dong1999, Takenaka2005}. Please note that the phenomenological descriptions of the optical conductivity does not rely on any specific nature of the localization, but generally discussed in terms of disorder effects. Here, we employed the model proposed by Fratini {\it et al.}, where the low-energy Drude-response is modified with the electron backscattering \cite{Fratini2014, *Fratini2016, *Fratini2020}, while the high-energy tail of the modified Drude response is still defined with the elastic scattering rate of the free carriers. These optical fingerprints commonly go hand in hand with a linear-in-$T$ resistivity (so-called bad metallic behavior), where the shift of the peak scales with the dc-conductivity \cite{Luca2017}.

To better demonstrate the phenomenological description of the peak in Fig.~\ref{Magnetic}(d), we subtract the response of the Dirac bands, as well as the van-Hove singularities from the measured optical conductivity. The solid lines are the best fits to the model suggested by Fratini {\it et al.} \cite{Fratini2014}. The high-temperature data are well reproduced by taking into account the backscattering correction, where we restrict ourselves with the static limit. We estimate the backscattering time of the electrons, $\tau_b$, to be around 13~fs at room temperature and increases to 76~fs at $T=150$~K. Below the transition at 150~K, the description breaks down: the observed feature corresponds to a very sharp Fano-like peak rather than a modified Drude-peak. This is in accord with the breakdown of the scaling relation between the dc conductivity and the linear shift of the peak energy below 150~K, displayed in Fig.~\ref{Magnetic}(e). Indeed, the presence of spin reorientation calls for a different approach at low temperatures. A better accuracy can be satisfied by taking into account a coupling of a Fano resonance to the electronic background. The Fano resonance starts to appear below $\sim$~150~K and gets more pronounced with decreasing temperature. Along with a stronger coupling to the electronic background, the behavior of the Fano-resonance follows the magnetization of the sample and the spin reorientation.

The appearance of such a sharp peak resembles excitations between electronic bands at $E_F$; i.e.\ it suggests the development of a partial gap in the density of states. This can be explained by additional Weyl nodes recently predicted for Fe$_3$Sn$_2$\ \cite{Yao2018}. Their existence is linked to the spin directions of the iron atoms within the kagome plane. When the spins reorient within the plane, the Weyl nodes become gapped, for the certain direction of magnetization, while for the other in-plane direction, there should be no gap. Hence, we conclude that the spin reorientation to the in-plane direction opens up the gap at the Weyl points, that we detect by optical means. The gap energy estimated from our measurements is 98~cm$^{-1}$\ (12 meV), which is in accord with the gap energy expected from calculations \cite{Yao2018}.

It is not surprising that these Weyl nodes are not identified by ARPES, because the current optical measurements possess a much higher energy resolution. Moreover, since the magnetic properties and the electronic structure are very sensitive to small fields, signatures of the gap might be missed in the magneto-transport measurements. The current optical study was conducted in zero field, taking into account only the inherent magnetization of the compound. This allows us to discover experimentally the proposed gap opening.

In summary, temperature and frequency-dependent infrared studies on the magnetic kagome metal Fe$_3$Sn$_2$\  reveal optical fingerprints of strongly correlated flat bands and topologically nontrivial Dirac fermions. The two-step absorption feature that evolves in the frequency-independent optical conductivity indicates the existence of two-dimensional Dirac cones  shifted in energy with respect to each other. The flat bands are seen as multiple absorption peaks in the low-energy optical conductivity. One of the peaks exhibits a strong shift in energy. At high temperatures, this peak reveals striking similarities with the displaced Drude-peak as an indication of localization effects. Below the spin reorientation temperature around 120~K, a gap opens around the Fermi energy as signature for theoretically proposed gapped Weyl nodes. On the other hand, spin reorientation seems not to effect the Dirac nodes buried well into the Fermi energy.

\begin{acknowledgments}
Authors acknowledge the fruitful discussions with L. Z. Maulana and A.V. Pronin and the technical support by G. Untereiner. H.C.L. acknowledges support from the National Key R\&D Program of China (Grants No. 2016YFA0300504, 2018YFE0202600), and the National Natural Science Foundation of China (No. 11574394, 11774423, 11822412). The work has been supported by the Deutsche Forschungsgemeinschaft (DFG) via DR228/48-1 and DR228/51-1. E.U. acknowledges the European Social Fund and the Baden-W\"{u}rttemberg Stiftung for the financial support of this research project by the Eliteprogramme.
\end{acknowledgments}

\bibliography{Fe3Sn2references}

\begin{thebibliography}{50}
\bibitem [{1()}]{1}%
  \BibitemOpen
  \href@noop {} {}\bibinfo {note} {It is interesting to examine the scattering
  rate of this sharp Drude-component. As the momentum-relaxing scattering of
  the free carriers (in the current case is the Dirac fermions as explained in
  the text) can be observed in the optical conductivity. Hence, one can
  directly obtain the corresponding scattering time, $\tau $, from the real
  part of the optical conductivity, $\sigma _1(\omega )$. The simultaneous fits
  of the reflectivity and the optical conductivity of our low energy
  Drude-component reveals a scattering rate [$\gamma = 1/(2\pi c\tau )$] of
  2~cm$^{-1}$\ at 7~K, while it increases to $\sim $17~cm$^{-1}$\ at room
  temperature. These corresponds to the scattering times $\tau _ {7K} = 2.5$~ps
  and $\tau _ {300K} = 0.3$~ps. Considering the reported average Fermi velocity
  for Fe$_3$Sn$_2$, $v_F = 2.2\times 10^5$~m/s, the momentum relaxation lengths
  [$\ell = v_F\tau $] are calculated as $\ell _ {7K} = 0.55$~$\mu $m and $\ell
  _ {300K} = 66$~nm. Considering that the localized carrier response of the
  flat bands are separated in energy and does not contribute to the low energy
  dynamics (within the obtained scattering scales), these length scales suggest
  that Fe$_3$Sn$_2$\ readily might be a suitable platform to realize a
  ballistic transport.}\BibitemShut {Stop}%
\end{thebibliography}


\newpage
\widetext
\clearpage
\begin{center}
	\textbf{\Large Supplemental Material for}
\end{center}

\begin{center}
\textbf{\large``Spin-reorientation-induced band gap in Fe$_3$Sn$_2$: Optical signatures of Weyl nodes"\\}

\end{center}

\title{\textbf{\large``Spin-reorientation-induced band gap in Fe$_3$Sn$_2$: Optical signatures of Weyl nodes"\\}}

\begin{center}
A. Biswas$^1$, O. Iakutkina$^1$, Q. Wang$^2$, H. C. Lei$^{2,\color{red}*}$, M. Dressel$^1$, and E. Uykur$^{1,\color{red}\dag}$
\end{center}


\subsection{Samples}
Single crystals of Fe$_3$Sn$_2$\ were grown using self-flux method as described elsewhere \cite{Wang2016}. An as-grown sample with a good surface quality was chosen for the optical spectroscopy study. The (001)-plane lateral dimensions of the sample used in the infrared spectroscopy study is 1000~$\mu$m$ \times $800~$\mu$m$ \times$ 200~$\mu$m.

\subsection{Transport and magnetic measurements}
The temperature-dependent dc electrical resistivity of Fe$_3$Sn$_2$\ single crystals was measured with a home-built system. Measurements are carried out within the (001)-plane in four contact geometry. The experiments were performed on the same piece used for the optical investigations. The RRR ratio is determined as R$_{300K}$/R$_{4K}$ is 39, indicating the good quality of the sample. 

The H$\parallel$(001)-plane magnetic properties of the sample were measured in a magnetic property measurement system (Quantum Design MPMS). DC magnetic susceptibility in zero field cooling (ZFC) and field cooling (FC) configuration have been obtained with $\mu_0$H = 0.1~T field. Field-dependent magnetization measurements have been performed for various temperatures up to 2~T. The slope of the magnetization is calculated at low fields ranging from 0.05~T at 2~K to 0.2~T at 300~K.

\subsection{Infrared measurements}

\begin{figure}[b]
\centering
\includegraphics[width=1\linewidth]{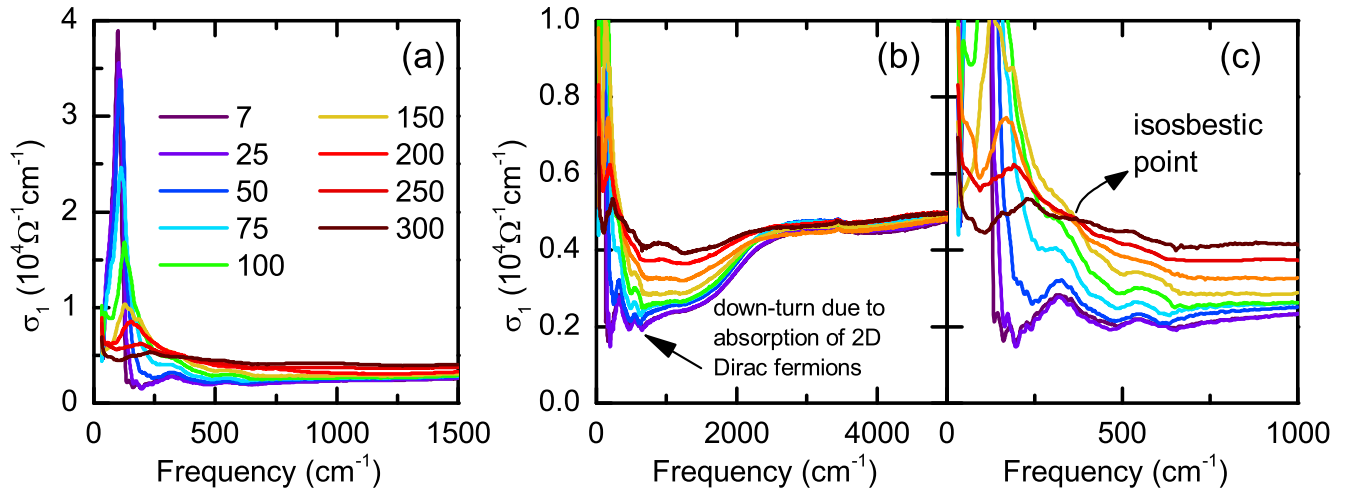}%
\caption{The real part of the optical conductivity for selected temperatures in different energy scales. (a) highlights the localization peak shifting to the lower energy range with decreasing temperature and evolving to the Fano-like peak below the spin reorientation temperature. (b) clearly demonstrate the two-step MIR absorption due to 2D Dirac fermions. (c) depicts the spectral weight transfer range. Down to 150~K, an isosbestic point reflecting the temperature-driven spectral weight transfer, changing its characteristics with spin-reorientation and strongly shifts to the lower energy range. }%
\label{OC_linear}%
\end{figure}

Temperature-dependent ($7~{\rm K} < T < 300$~K) optical reflectivity was measured on thin platelet-like as-grown crystal with the (001)-plane, covering a broad frequency range. The low-energy experiments (30-800~cm$^{-1}$) were conducted with a Bruker IFS 113v Fourier-transform infrared spectrometer; the reflectivity was obtained with an {\it in situ} gold overcoating technique. In the higher energy range up to 2~eV, we employed a Hyperion IR microscope coupled to a Bruker Vertex 80v spectrometer. In this energy range, the infrared beam is focused to around 100~$\mu$m and the freshly evaporated mirrors have been used as a reference. 

The optical conductivity is calculated via standard Kramers-Kronig analysis. Considering the highly metallic nature of the sample, we used a Hagen-Rubens extrapolation in the low-energy range. The obtained dc conductivity values agrees with transport measurements performed on the same sample [inset of Fig.1(b) of the main text]. For the high energy extrapolations, we used x-ray scattering functions \cite{Tanner2015}. The optical conductivity for selected temperatures have been given in linear scale in Fig.~\ref{OC_linear}(a) and (b) for the different energy ranges.

The skin depth of the infrared radiation used in the measurements exceeds 30~nm for all temperatures and frequencies (in the far-infrared range, it is above 150~nm). Hence, our optical measurements reflect the bulk properties of Fe$_3$Sn$_2$.

\subsection{Localized carriers and Dirac fermions}
In the real part of the optical conductivity, the strong absorption feature dominating the low energy spectra masks the Drude-like free carrier contribution, while the dc transport measurements corroborate the existence. In the measurement window employed here, we can only observe the small upturn of this Drude-component. Here we also examined other optical variables to ensure the existence of this sharp Drude-component, namely the imaginary part of the optical conductivity, $\sigma_2(\omega)$, where one can see the signatures of the low-energy Drude-contribution more clearly \cite{Dressel}. 

\begin{figure}[b]
\centering
\includegraphics[width=1\linewidth]{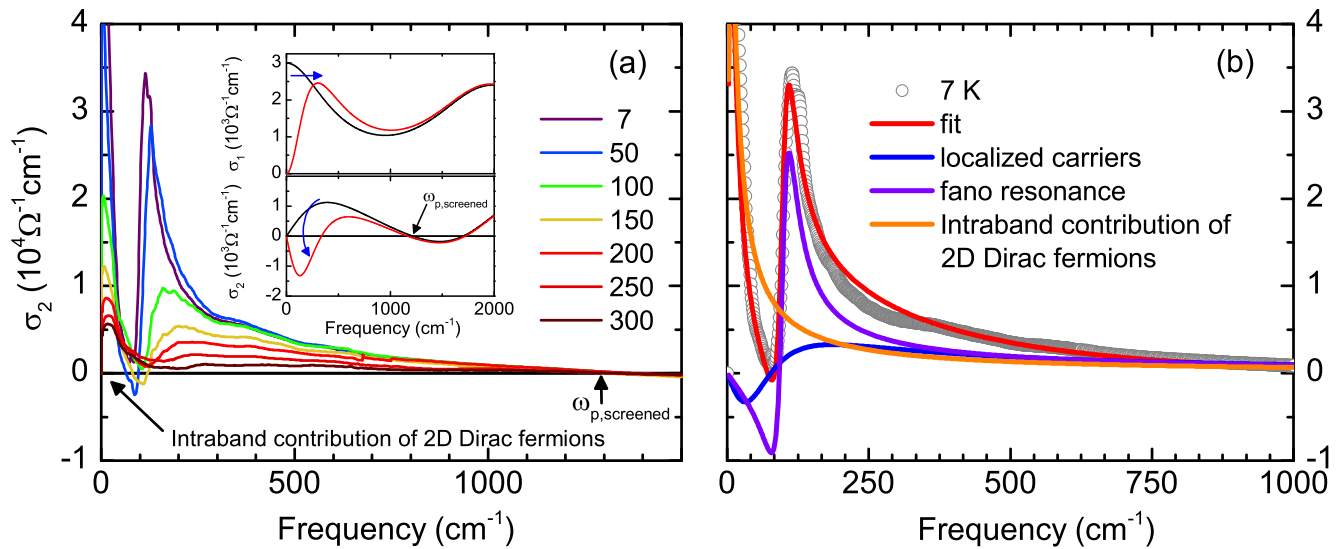}%
\caption{(a) Imaginary part of the optical conductivity for selected temperatures. The zero-crossing indicates the screened plasma frequency of the localized carriers, where a second contribution reflecting the intraband transitions of the 2D Dirac fermions is visible in the low energy range. Inset demonstrates the response of conventional Drude and modified Drude contributions to the real and imaginary part of the optical conductivity. (b) The decomposition of the imaginary part of the optical conductivity for 7~K spectrum as an example, where one can clearly distinguish the different contributions of the localized and free carriers along with the sharp Fano resonance.}%
\label{s2}%
\end{figure}

In Fig.~\ref{s2}(a), the imaginary part of the optical conductivity is given. The zero crossing of this optical variable is defined as the screened plasma frequency. For Fe$_3$Sn$_2$, we estimated the screened plasma frequency as $\sim$1300~cm$^{-1}$ and the temperature-dependence is negligibly small. The low-energy features reveals the signatures of the second plasma edge that can be associated with the free carrier response of the Dirac bands. As demonstrated in the inset of Fig.~\ref{s2}(a), in the case of Drude-like free carrier response with a high-energy absorption, the imaginary optical conductivity shows the zero-crossing at the screened plasma frequency followed by the broad maximum extrapolating to zero at $\omega \rightarrow 0$. In the case of the localized carriers, the Drude-peak is shifting to the finite frequencies and in return low-energy response of $\sigma_2$ is suppressed below zero. 

For the Fe$_3$Sn$_2$\ system, we can reproduce the spectra with a similar analogy; however, the low-energy response reveals an additional zero-crossing below the energies described for the localized carriers along with the maximum as in the case of a regular Drude-component indicating the second contribution at lower energies. Below the spin-reorientation temperature a sharp Fano resonance appear in the spectra indicating a strong coupling to the electronic background. 

We also want to point out that the zero-crossing of the imaginary optical conductivity is determined as the screened plasma frequency in the framework of the Drude interpretation, and in principle can be different for the Dirac fermions and/or localized carriers. On the other hand proposed models still consider it similar to the Drude approximation; therefore, here we used the same term.

\subsection{Decomposition of the optical conductivity}

\begin{figure}[b]
\centering
\includegraphics[width=1\linewidth]{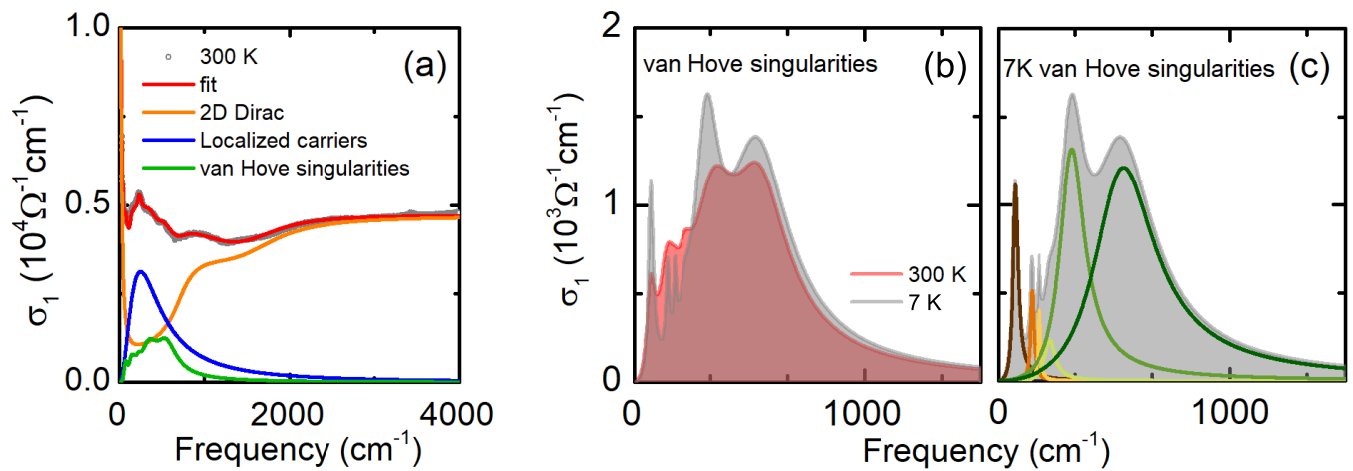}%
\caption{(a) The decomposition of the optical conductivity at room temperature. The 2D Dirac response dominates the high energy spectrum, while the lower energy part can be described with the modified Drude component, reflecting the localized carrier response and several absorption features identified as the van Hove singularities. (b) The temperature, evolution of van Hove singularities. The peak positions does not change with temperature, moreover, the overall spectral weight does not change significantly, either. On the other hand, modes getting sharper, with decreasing temperature, as expected. (c) The decomposition of the low-energy absorption feature at 7~K. We identified 6 lorentzian mode to describe our spectra throughout the measured temperature range placed at 72, 141, 172, 223, 311, 542~cm$^{-1}$. }%
\label{decompositions}%
\end{figure}

We determined that the optical conductivity spectra can be decomposed into two main parts, namely the response of the Dirac fermions and the localized carriers of the flat bands. We can further elaborate these contributions as the inter- and intra-band contributions of the 2D-Dirac fermions, conduction response of the localized carriers (modified Drude-component), and the temperature-independent van Hove singularities. At lower temperatures (below the spin-reorientation temperature), the modified Drude component alone cannot describe the strong low energy absorption. Instead, we defined this absorption with a Fano resonance \cite{Fano1961}, as described in Eq.~\ref{fano}, that is coupled to the localized carriers. In Fig.~\ref{decompositions} a sample fit to the room temperature spectrum has been given. The fit to the low temperature spectrum with the strong absorption feature has been shown in the main text, Fig.2(a).

\begin{equation}
\sigma_1(\omega) = \frac{2\pi}{Z_0}\frac{\Omega^2}{\gamma}\frac{1+\frac{4q(\omega-\omega_0)}{\gamma}-\frac{1}{q^2}}{1+\frac{4(\omega-\omega_0)^2}{\gamma^2}}
\label{fano}
\end{equation}

\begin{figure}[h]
\centering
\includegraphics[width=1\linewidth]{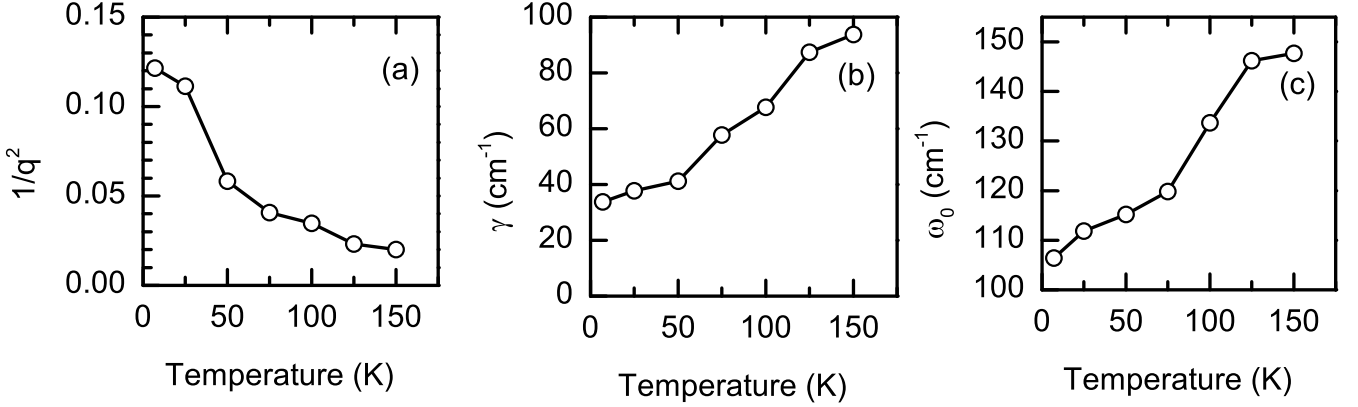}%
\caption{The fit parameters of the Fano resonance. (a) dimensionless asymmetry parameter, (b) the linewidth of the resonance, and (c) the resonance frequency of the mode. }%
\label{fanopara}%
\end{figure}

Here, $Z_0$ is the vacuum impedance, $\omega_0$, $\gamma$, and $\Omega$ is the resonance frequency, line width, and the strength of the Fano mode, and $q$ is the dimensionless parameter that describes the asymmetry of the Fano resonance. Larger 1/$q^2$ shows a stronger asymmetry, while for 1/$q^2$=0, the regular Lorentzian line shape is recovered. Below the spin-reorientation temperature (around 150~K), this mode start to be prominent in the spectra and shows a strong temperature dependence. In Fig.~\ref{fanopara}, the fit parameter to this mode is given. The asymmetry of the mode (Fig.~\ref{fanopara}(a)) shows a very strong temperature dependence indicating a strong coupling to the electronic background, triggered with the spin reorientation. The Fano mode gets sharper and the resonance frequency shows a red shift with decreasing temperature; however, as demonstrated in Fig.3(c) of the main text, this shift in energy is much smaller than what is expected for the localization peak.  

As several flat bands have been determined in the vicinity of the Fermi energy, several absorption features are also expected in the optical conductivity spectra, van Hove singularities arising due to the transitions between flat bands. Indeed several temperature-independent absorption modes are visible in the optical conductivity spectra, where the resonance frequency does not change, while the modes getting sharper. Although, we cannot discard the existence of more, we determined six of these van Hove singularities at resonance energies: 72, 141, 172, 223, 311, 542~cm$^{-1}$. In Fig.~\ref{decompositions} the decomposition of these bands for 7~K spectrum have been given along with the temperature evolution in Fig.~\ref{decompositions} (b). Please note that the energy range where these van Hove singularities are dominant mostly covers the energies below the absorptions of 2D Dirac fermions, and the signature of the downturn due to step-like behavior is also clearly visible in the spectra (Fig.~\ref{OC_linear} (b)). 

\subsection{Spectral Weight Analysis}

In Fig.2 of the main text, a spectral weight (SW) analysis is shown for the different contributions to the optical conductivity. SW of the spectra have been calculated as 

\begin{equation}
SW = \int_{0}^{\omega_{cut-off}} \sigma_1(\omega) d\omega
\end{equation}

\begin{figure}[t]
\centering
\includegraphics[width=1\linewidth]{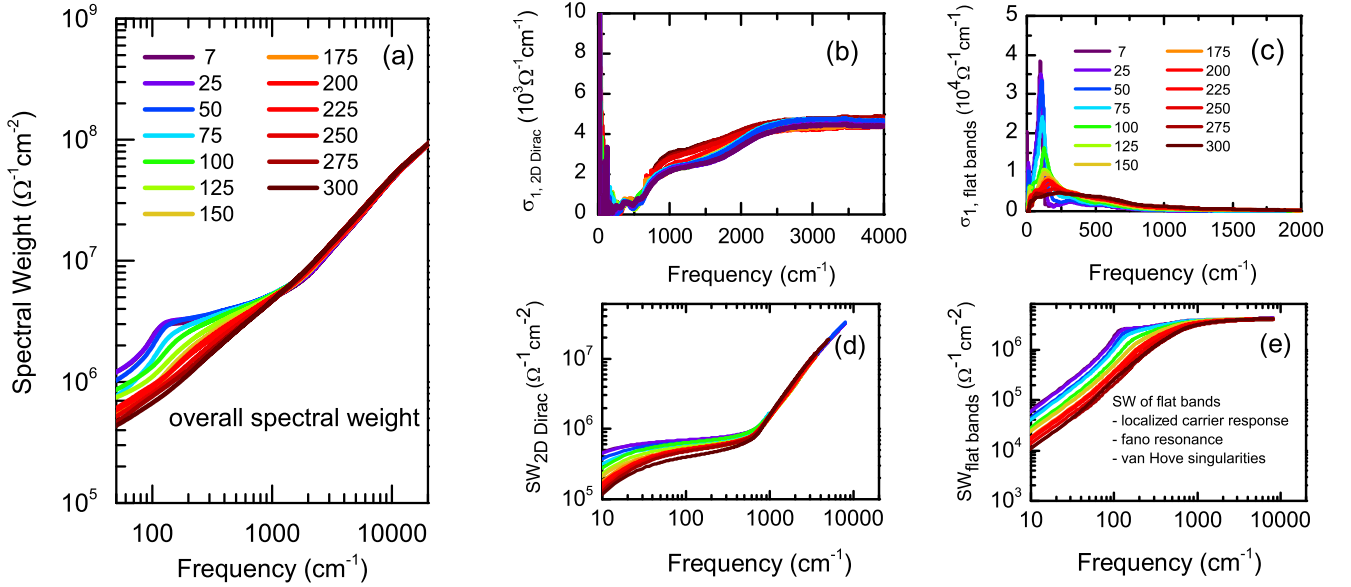}%
\caption{(a) The temperature- and frequency-dependent overall spectral weight. The spectral weight transfer is completed below $\sim$~1~eV. (b) 2D Dirac response after the low-lying absorption features are subtracted. (c) Low lying absorption features, namely, the localized carriers, van Hove singularities, and the strong Fano resonance at low temperatures. (d) and (e) is the temperature- and frequency-dependent spectral weight of the individual contributions respectively. }%
\label{SW}%
\end{figure}

For the full spectra, the cut-off frequency have been chosen as 2~eV that reflects the whole measurement range. Although the overall SW transfer happens within 1~eV. The energy-dependent spectral weight for several temperatures have been given in Fig.~\ref{SW} (a), demonstrating the conservation of the spectral weight and the energy range of the changes. 

For the individual contributions, we first fit the whole spectrum with Drude (intraband contribution of Dirac fermions), Lorentzians (for the van Hove singularities), modified Drude component (for the localized carriers), and the Dirac contributions (reflecting the frequency-independent contributions). An example fit is given in Fig.~\ref{decompositions} (a) The dc-conductivity estimated from the optical measurements have been kept as the dc-limit during the fits of the spectra. For the Dirac SW (interband and intraband), the low-lying absorption bands have been subtracted and the remaining spectra were integrated up to 1~eV. For the flat-band SW contribution, we followed the opposite trend, namely, we subtracted the Dirac contributions (intraband and interband) and integrated the spectra up to 1~eV. In Fig.~\ref{SW} (a) and (b), the remaining spectra after the subtraction process used for the spectral weight analysis have been given. The SW conservation for the individual contributions holds as in the case of overall SW. The energy-dependent spectral weight for several temperatures have been given in Fig.~\ref{SW}, demonstrating the conservation of the spectral weight for the individual contributions and the energy range of the changes. One can also realize that the described Dirac points does not change significantly with temperature, while temperature smear out the step-like features.

\begin{figure}[b]
\centering
\includegraphics[width=1\linewidth]{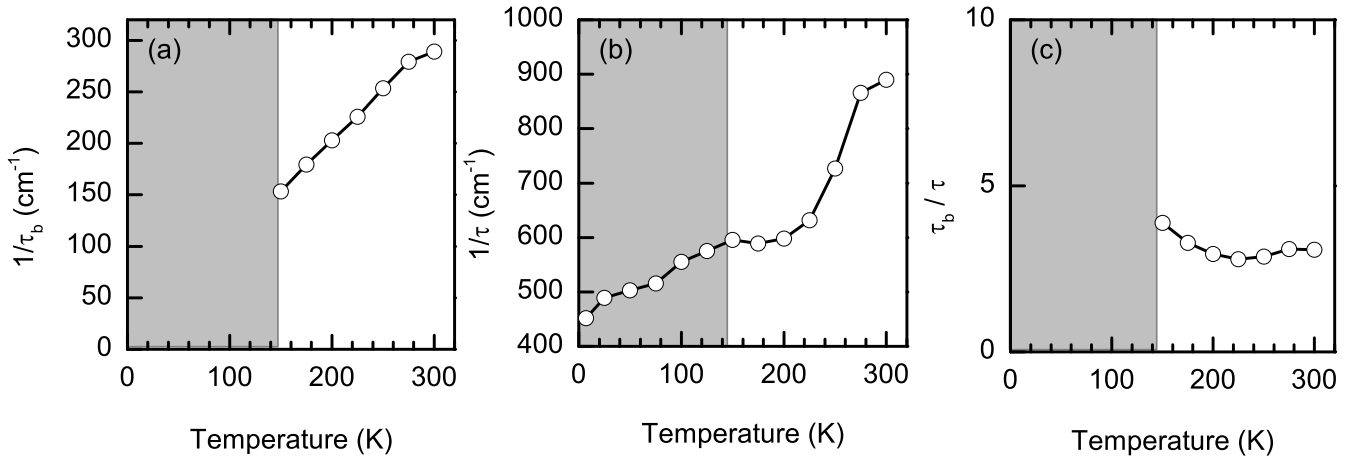}%
\caption{Fit parameters of the model based on Eq.~\ref{Fratini}. (a) Backscattering of the electrons, (b) elastic scattering time, and (c) is the ratio between two time scales.  }%
\label{Fratinipara}%
\end{figure}

\subsection{Optical conductivity of localized carriers}
The optical conductivity that has been discussed in terms of localized carriers in the main text have been analysed with the model proposed by Fratini $et$ $al.$ as described in \cite{Fratini2014, Fratini2016, Fratini2020}. Our description is restricted within the static limit, considering the limited low-energy measurement range. The inelastic scattering parameter within the static limit goes to infinity. So, within static limit, in Eq \ref{Fratini} $\tau_{in} \rightarrow \infty$ for real part of the optical conductivity.

\begin{equation}
\sigma_1(\omega) = \frac{C}{\tau_b-\tau}\frac{\text{tanh}(\frac{\hbar\omega}{2k_BT})}{\hbar\omega}\times \text{Re}\left[\frac{1}{1+\frac{\tau}{\tau_{in}}-i\omega\tau}-\frac{1}{1+\frac{\tau_b}{\tau_{in}}-i\omega\tau_b}\right]
\label{Fratini}
\end{equation} 

This equation describes the modified Drude component mentioned in the main text, where the position of the localization peak is determined by the backscattering rate of the electrons (1/$\tau_b$) and the high-frequency tail is controlled by the elastic scattering rate (1/$\tau_{in}$). In Fig.~\ref{Fratinipara}, the obtained fit parameters have been given as a function of temperature. Below the spin-reorientation temperature (marked with the gray area in Fig.~\ref{Fratinipara}), the backscattering of the electrons cannot be determined accurately due to the strong influence of the Fano resonance. Therefore, we remove it from the discussion, but the elastic scattering part can be determined from the imaginary part of the optical conductivity, as demonstrated in Fig.~\ref{s2} (b).

\end{document}